\documentclass[aps,showpacs,groupedaddress]{revtex4}  
\usepackage{graphicx}  
\usepackage{dcolumn}   
\usepackage{bm}        
\usepackage{amssymb}   

\begin{document}

\hspace{4.2in} \mbox{FERMILAB-CONF-08-277-E}
\vspace{1cm} 

\title{Measurements of $t\bar{t}$ production cross-section with D0 experiment}
\author{V.~Shary for D0 collaboration}
\altaffiliation{CEA, IRFU, SPP
Centre de Saclay, F-911191 Gif-sur-Yvette, France}

\begin{abstract}

The recent measurements of the top anti-top quark pair production cross-section in proton antiproton 
collisions at $\sqrt{s}=1.96$ TeV in lepton + jets and  dilepton (including tau lepton)  channels are presented.
These measurements are based 
on  1~fb$^{-1}$ of data collected with the DO experiment at 
the Fermilab Tevatron collider. 
The measured values are compatible with the standard model prediction and have the uncertainty
close to the uncertainty of the theoretical prediction $\sim 10 \%$.
\end{abstract}


\maketitle

\section{Introduction}

In the standard model framework 
the production cross-section of the top quark  top antiquark pair ($t\bar{t}$) in proton-antiproton collisions could be calculated
in the perturbative QCD approach.
The most recent calculations are performed at the next-to-leading order  with the next-to-next-leading order soft correction
 (NLO+NNLO)~\cite{Kidonakis:2008mu} or at the next-to-leading order with the next-to-leading threshold  logarithm correction 
(NLO+NLL)~\cite{Cacciari:2008zb}.
The exact value of the calculated cross-section depends on the top quark mass ($m_t$)
and the choice of parton distribution functions (PDFs).
For example, choosing CTEQ6.6 parametrization~\cite{Nadolsky:2008zw}
 leads to a value of $7.39^{+0.57}_{-0.52}$ pb  (for $m_t=172$~GeV) for~\cite{Kidonakis:2008mu}. 
In~\cite{Cacciari:2008zb} for CTEQ6.5 PDFs~\cite{Tung:2006tb} the cross-section value is calculated to be  
$7.61^{+0.30}_{-0.53}$ (scales) $^{+0.53}_{-0.36}$ (PDFs) pb (for $m_t=171$~GeV).
In both cases uncertainties on the calculated value have two main contributions: uncertainties from the choice of 
PDF and  the renormalization and factorization scales, set to be both equal to $m_t$. The scales uncertainty
is determined by varying both scales from $m_t/2$ to $2m_t$. 
These scales could be varied simultaneously or independently,
 and different approaches may affect a lot the scale uncertainty~\cite{Cacciari:2008zb}.
Despite some what different approaches,  both groups~\cite{Kidonakis:2008mu} 
and~\cite{Cacciari:2008zb} give compatible values for the $t\bar{t}$
cross-section, which are also compatible between different sets of PDFs. The uncertainties on these 
theoretical predictions are estimated to be less than~8\%.

In this overview we report the $t\bar{t}$ production cross-section measurements based on the data collected by
 the D0 
detector between  August 2002 and February 2006 with an integrated luminosity of about~1 fb$^{-1}$.
The description of the D0 detector can be found elsewhere~\cite{Abazov:2005pn}.
In the standard model the top quark decays to a $W$ boson and a $b$ quark  with a probability close to~100\%. 
Here we consider the final state where one $W$ boson decays to a lepton and another one to quarks (lepton + jets final state, 
branching ratio 38\%) 
and the final state where both $W$ bosons decay to leptons (dilepton final state). For the dilepton final state we consider separately
final states with one hadronically decaying tau lepton (3.6\%) and final states which contain electron and muon only the latter
includes $\tau\to e,\mu$ decays~(6.5\%).  

\section{Electron, muon dilepton final states}
In order to maximize the  acceptance we consider the following final states:  the ones with well identified
electron or muon ($ee$, $e\mu$, $\mu\mu$),  and the final states with one well identified 
electron or muon and the reconstructed 
track ($e$ + track , $\mu$+ track). Selection criteria are optimized in each channel separately  to yield the best possible precision.
Possible overlaps between different channels are removed by applying veto cuts. The typical identification criteria for an electron
require an isolated cluster in the electromagnetic part of the calorimeter with a reconstructed transverse momentum $p_T \ge 15$~GeV.
Electron must be matched to a reconstructed track and located in the  region 
$0 \le |\eta|\le 1.1$ or $1.5 \le |\eta| \le 2.5$, where $\eta$ is a pseudorapidity defined as $\eta = - ln(\tan(\theta/2))$
 and $\theta$ is the polar angle with the proton beam. Muon is reconstructed as a track in the muon system, matched to a track
from the tracking detector, isolated from the other objects  in the calorimeter and in the tracking detector. Muons are required
to be within $|\eta|\le 2.0$ and have $p_T \ge 15$~GeV.  
Two $b$ quarks from top quark decays reconstructed as jets in the calorimeter using iterative  cone algorithm~\cite{Blazey:2000qt} with a cone radius of 0.5. Jets should be within $|\eta| \le 2.5$ and typically are required to have $p_T \ge 20$~GeV, where 
$p_T$ is corrected for the jet energy scale including the correction for muons from semileptonic $b$ quark decays.
In all channels except $e\mu$ a cut on the missing transverse energy corrected for the  jet energy scale and muons $p_T$ allows to 
improve signal-to-background ratio due to the presence of neutrinos from $W$ boson decays. 
The typical cut value is 20 -- 40~GeV.
Additional topological selections exploiting the differences in the signal and background kinematics are  applied in each channel.
Contribution of the main 
backgrounds,  $Z$ boson events decaying to leptons and diboson events from $WW$, $WZ$, $ZZ$ production
are estimated from MC simulation using Alpgen or Pythia generators. 
Data events are used to estimate instrumental backgrounds
originating from jets misidentified as electrons or muons from the semileptonic $b$ quark decays and events with large missing 
transverse energy due to the detector resolution effects. 
Table~\ref{tab:dilepton} summarizes the observed number of events, expected background and efficiency for the 
$ee$, $e\mu$, $\mu\mu$ channels.
This allows to measure the $t\bar{t}$  cross-section with 22\% precision: 
\[
\sigma = 6.8^{+1.2}_{-1.1} \ (stat) ^{+0.9}_{-0.8} \ (syst) \pm 0.4 \ (lumi) \ pb \quad (m_t = 175~GeV)
\]

\begin{figure}[!hbt]
\centerline{\includegraphics[width=.4\textwidth]{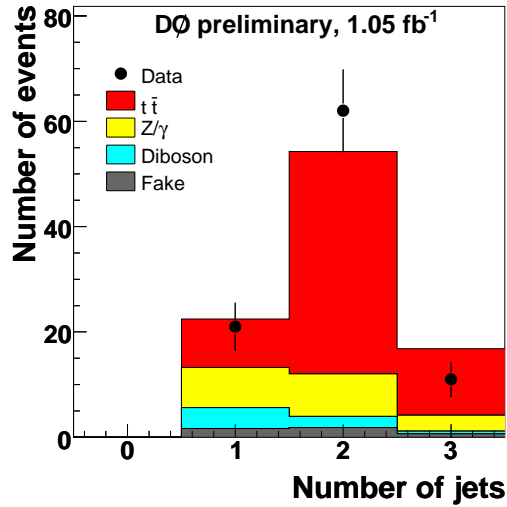}}
\caption{Jet multiplicity spectra for dilepton and 
lepton+track channels. The last bin is inclusive.
\label{fig:njets_dilepton}
}
\end{figure}

In the lepton + track channel the signal-to-background  ratio is much smaller than in other dilepton channels. In order to improve
it an additional requirement that at least one jet is identified as a b quark  jet (b-tagging) is applied. The b-jet identification 
at D0 is based on the neural network, which combines several parameters sensitive to the displaced decay vertices of the B hadrons.  
A typical cut on the neural network output used by the top analysis allow to tag b quark jets with an efficiency near 54\% 
and the mistagging rate (probability to tag a light quark jet) near 1\%. 
Table~\ref{tab:ltrack} summarizes the observed number of events, expected background and efficiency for the 
$e$~+~track and $\mu$~+~track  final states after applying veto selection on  $ee$, $e\mu$, $\mu\mu$.
The jet multiplicity spectra for $t\bar{t}$ signal and backgrounds events are shown on fig.~\ref{fig:njets_dilepton}.
Combination of  $ee$, $e\mu$, $\mu\mu$, $e$~+~track and $\mu$~+~track final states  allows to increase the precision of the
$t\bar{t}$ cross-section measurement to  19\%:
\[
\sigma = 6.2 \pm 0.9  \ (stat) ^{+0.8}_{-0.7} \ (syst) \pm 0.4 \ (lumi) \ pb\quad (m_t = 175~GeV)
\]
Systematics uncertainties in dilepton channels come from several sources: jet energy calibration,
identification of jets, muons, electrons and tracks,  trigger efficiency, instrumental background contribution,
background normalization,  b-tagging efficiency (for lepton+track only). Many of these sources contribute 
at the same level and hence there is no ``main'' systematic uncertainty which can be reduce easily. Instead  the laborious
work on each source of systematics is required for further improvement.

\begin{table}
\begin{tabular}[t]{l|ccccc}
\hline 
Channel & Observed & $N^{bkg}$ & BR& L(pb$^{-1}$) & $\varepsilon$ \\ \hline
$ee$ & 16 & 3.0 & 0.01584 & 1036 & 8.3 \% \\
$e\mu$ nj=1 & 16 & 10.2 & 0.03155 & 1046 & 3.1 \% \\
$e\mu$ nj=2 & 32 & 6.7 & 0.03155 & 1046 & 12.4 \% \\
$\mu\mu$ & 9 & 3.6 & 0.01571 & 1046 & 5.1 \% \\
\hline 
\end{tabular}
\caption{\label{tab:dilepton} 
Observed event yield, expected background, branching ratio, integrated luminosity and efficiency 
for  dilepton $ee$, $e\mu$ (1 and 2 jets selections), $\mu\mu$ channels.} 
\end{table}

\begin{table}
\begin{tabular}[t]{l|ccccc}
\hline 
Channel & Observed & $N^{bkg}$ & BR& L(pb$^{-1}$) & $\varepsilon$ \\ \hline
$e$+track nj=1& 14& 1.58 & 0.1066 & 1035 & 0.25 \% \\
$e$+track nj $\ge 2$ & 8 & 1.83 & 0.1066 & 1035 & 1.31 \% \\
$\mu$+track nj=1 & 1 & 1.38 & 0.1066 & 994 & 0.18 \% \\
$\mu$+track nj $\ge 2$& 8 & 1.36 & 0.1066 & 994 & 1.08 \% \\
\hline 
\end{tabular}
\caption{\label{tab:ltrack} 
Observed event yield, expected background, branching ratio, integrated luminosity and efficiency 
for $e$+track and $\mu$ + track channels  with 1 and 2 jets selections.} 
\end{table}

\section{Dilepton final states with hadronically decaying tau lepton} 

Tau lepton is the heaviest lepton and has  the strongest 
coupling to the Higgs boson. Final states with one hadronically decaying tau lepton 
contains much more background than other dilepton final states, mainly from multijet and $W$ + jets events, but 
it is interesting to study them because they are  sensitive to the contribution from the 
physics beyond the  standard model. For example, if the charged Higgs boson in the minimal supersymmetric standard model
scenario has a low enough mass, the $t\to H^{+}b$ decays will be allowed and will enhance the tau lepton final states.

The tau lepton is identified  as a narrow jet in the calorimeter with a cone size 0.3
matched to one or more tracks. The neural network (NN) is used to 
distinguish tau leptons and jets. The NN input parameters describe energy deposition profile in the calorimeter 
 and  track-calorimeter correlations.  NN is trained on the MC $Z\to\tau\tau$  sample
and on the background sample from real data.  The performance of neural network has been verified with $Z\to\tau\tau$ data.
For more details see~\cite{d0-tau}.
The event selection for $e\tau$ and $\mu\tau$ final states require at least one tau candidate and one electron 
with $p_T > 15 $~GeV or muon with $p_T > 20 $~GeV,  at least two jets with $p_T > 20$~GeV and leading jet $p_T > 30$~GeV, 
transverse missing energy between
15 and 200~GeV. Further improvement of the signal-to-background ratio is reached by the requirement that at least 
one jet is b-tagged. The main physics background, $W$+jets and $Z$+jets, is  estimated from MC, and multijet background
where jets are misidentified as  taus  is estimated from data.
The measured cross-section has a precision about 30\%, 
where all $t\bar{t}$ dilepton and lepton+jets final states considered as a signal:
\[
\sigma = 8.3 ^{+2.0}_{-1.8}  \ (stat) ^{+1.4}_{-1.2} \ (syst) \pm 0.5 \ (lumi) \ pb \quad (m_t = 175~GeV)
\]
To measure the cross-section  times branching ratio we consider only lepton + tau final state as a signal:
\[
\sigma(t\bar{t}) BR(t\bar{t}\to l \nu_l \tau \nu_\tau b\bar{b}) = 0.19 \pm 0.08 \ (stat) \pm 0.07 \ (syst) \pm 0.01 \ (lumi)\  pb
\]
which is compatible with a standard model prediction 0.128 for electron and 0.125 for muon  final states.

\section{Lepton + jets final states}

In lepton + jets final state one $W$ boson 
decays to a lepton (electron or muon) and another one to jets.  In this final state two approaches
have been used to separate signal from background: 
kinematic likelihood and b-tagging. In both cases, first we define an inclusive sample 
 selected by requiring exactly one isolated electron
or muon with
$p_T > 20$~GeV and $|\eta| \le 1.1$ for electron and $|\eta| \le 2.0$  for muons,
missing transverse momentum  $\ge 20$ (for $e$ + jets) or
$\ge 25$~GeV (for $\mu$ + jets) and at least three jets with $p_T \ge
20$~GeV and  $|\eta| \le 2.5$.  
The leading jet must have $p_T \ge 40$~GeV,
and the lepton $p_T$ and the missing transverse energy vector must be separated in
azimuth to reject background events with mismeasured
particles. 
At this stage the data sample contains only near 20\% of $t\bar{t}$ events.  The main background
originated  from $W$+jets events and multijet events.
$W$ + jets events contain real electrons or muons coming from $W$ decays.  In multijet events  one of the jets
is misreconstructed as an electron or produces a muon because of the semileptonic b quark decays.
We determine the multijet background contribution in the selected  sample  
by using data samples with  the relaxed electron
identification or the muon isolation requirements. 

In the kinematic likelihood analysis the further improvement of the signal-to-background ratio is reached 
by  using a kinematic determinant built with 
several variables exploiting the difference in kinematics between $t\bar{t}$ and backgrounds events.
For the events with exactly three jets we also use additional requirement that sum of jets transverse momenta is less than 120~GeV.
Simultaneous fit of $e$+jet and $\mu$+jet channels with exactly three jets or with four and more jets 
allows to extract the number of $t\bar{t}$ and background events.
For this we use the discriminant templates from MC for the $t\bar{t}$ signal and $W$+ jets background  and from data
for the multijet background. The sample compositions for event in the third  and forth jet bins is presented in tab.~\ref{tab:ljets}.
The measured cross-section value  has a 15\%  precision:
\[
\sigma = 6.6 \pm 0.8\  (stat) \pm 0.4\  (syst) \pm 0.4\ (lumi) \ pb \quad (m_t = 175~GeV)
\]

\begin{figure}[!hbt]
\includegraphics[width=.48\textwidth]{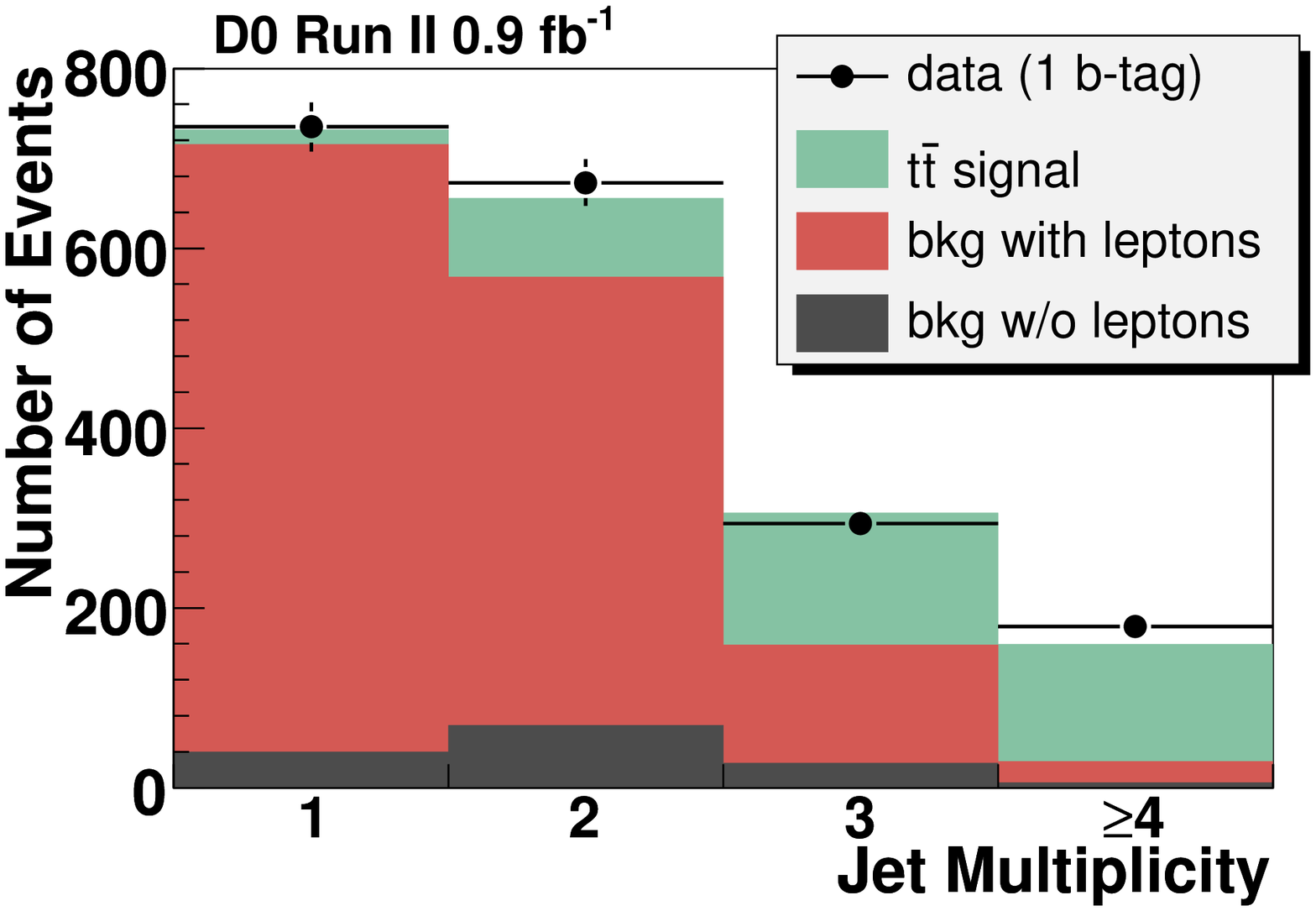}
\hfill
\includegraphics[width=.48\textwidth]{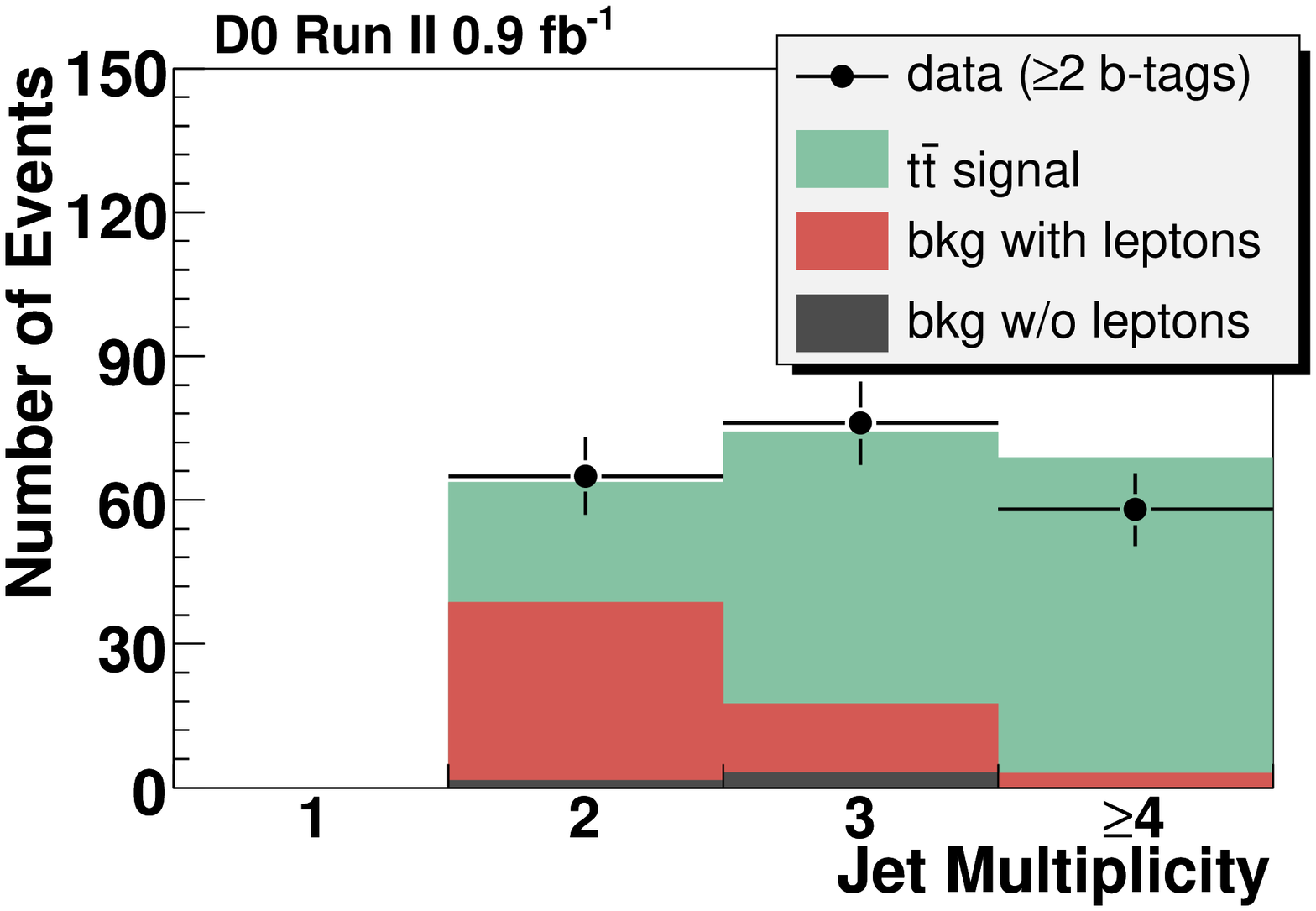}
\caption{Jet multiplicity spectra for leptons + jets events  with one b-tagged jet (left) and  with at least two
b-tagged jets (right). 
The histograms show (from top to bottom) the contributions from $t\bar{t}$ production, from backgrounds with
leptons, mainly $W$~+~jets, and from the multijet background.
\label{fig:njets_ljets}
}
\end{figure}

The b-tag analysis requires that at least one jet is b-tagged.
We determine the number of multijet events as above and
the number of events expected from other background
sources from the number of background events in the
inclusive sample multiplied by their probability to be b-tagged.
We obtain the b-tagging probability from the MC simulation
corrected for differences in the efficiencies
observed in the simulation and in data.  The composition
of the b-tagged samples is given in tab.~\ref{tab:ljets} .
The jet multiplicity spectra for $t\bar{t}$ signal and backgrounds events are shown on fig.~\ref{fig:njets_ljets}.
The calculated cross-section has a precision of 12\%: 
\[
\sigma = 8.1 \pm 0.5\  (stat) \pm 0.7\  (syst) \pm 0.5\ (lumi) \ pb \quad (m_t = 175~GeV)
\]

\begin{table}
\begin{minipage}[t]{.4\textwidth}
\begin{tabular}[t]{l|cc}
\hline 
 & 3 jets & $\ge$ 4 jets \\ \hline
$N_{data}$ & $1760$ & $626$ \\
$N_{t\bar{t}}$ & $245 \pm 20$ & $233\pm19$ \\
$N_{W+jets, other} $  & $ 1294 \pm 48 $ & $321 \pm 30 $ \\
$N_{multijet}$ & $ 227 \pm 28 $ & $ 70 \pm 12$ \\
\hline 
\end{tabular}
\end{minipage}
\hfill
\begin{minipage}[t]{.57\textwidth}
\begin{tabular}[t]{l|cccc}
\hline 
&\multicolumn{2}{c}{3 jets} &\multicolumn{2}{c}{ $\ge$ 4 jets} \\
 & 1tag & $\ge$2tags & 1tag &$\ge$2tags\\ \hline
$N_{t\bar{t}}$ & $147 \pm 12$ & $57\pm 6$ & $130 \pm 10$ & $66\pm 7$ \\
$N_{W+jets} $   & $105 \pm 5$ & $10\pm 1$ & $16 \pm 2$ & $2\pm 1$ \\
$N_{other} $  & $27 \pm 2$ & $5\pm 1$ & $8\pm 1$ & $2\pm 1$ \\
$N_{multijet}$ & $27 \pm 6$ & $3\pm 2$ & $6 \pm 3$ & $0\pm 2$ \\
Total &  $306\pm 14$ & $74\pm6$ & $159\pm11$ & $69\pm7$ \\ \ hline
$N_{data}$ & $294$ & $76$ & $179$ & $58$ \\
\hline 
\end{tabular}
\end{minipage}
\caption{\label{tab:ljets} Sample composition for lepton+jets final state in likelihood analysis (left) and 
b-tagging analysis (right). Number of $t\bar{t}$ events is calculated using the cross-section measured by the likelihood or 
by the b-tagging analyses correspondingly. }
\end{table}

The breakdown of systematics uncertainties is shown in tab.~\ref{tab:ljets_syst}.  
The b-tag analysis systematics is larger than the kinematic likelihood one, mainly because of the 
systematics on the b-tagging efficiency, but b-tag analysis has
better statistical uncertainty due to the better signal to background separation.
Combining  two results together allows
to improve overall precision, because these two results are only partially correlated (statistical correlation factor is 0.31).
This combined cross-section value  has the relative uncertainty 11\%. This is the most precise measurement of
the $t\bar{t}$ production cross-section at D0~\cite{Abazov:2008gc}.
\[
\sigma = 7.4 \pm 0.5\  (stat) \pm 0.5\  (syst) \pm 0.5\ (lumi) \ pb \quad (m_t = 175~GeV) .
\]

\begin{table}
\begin{tabular}[t]{l|ccc}
\hline 
Source &b-tag & Likelihood & Combined \\ \hline
Selection efficiency & 0.26 pb &0.25 pb &0.25 pb\\ 
Jet energy calibration &0.30 pb &0.11 pb &0.20 pb\\ 
b tagging &0.48 pb &- &0.24 pb\\ 
MC model &0.29 pb &0.11 pb &0.19 pb\\ 
Multijet bckg normalization &0.06 pb &0.10 pb &0.07 pb\\ 
Likelihood fit& - &0.15 pb &0.08 pb\\  \hline
\end{tabular}
\caption{\label{tab:ljets_syst} Breakdown of systematic uncertainties in lepton+jets channel.}
\end{table}

\section{Conclusion}

\begin{figure}[!hbt]
\centerline{\includegraphics[width=.7\textwidth]{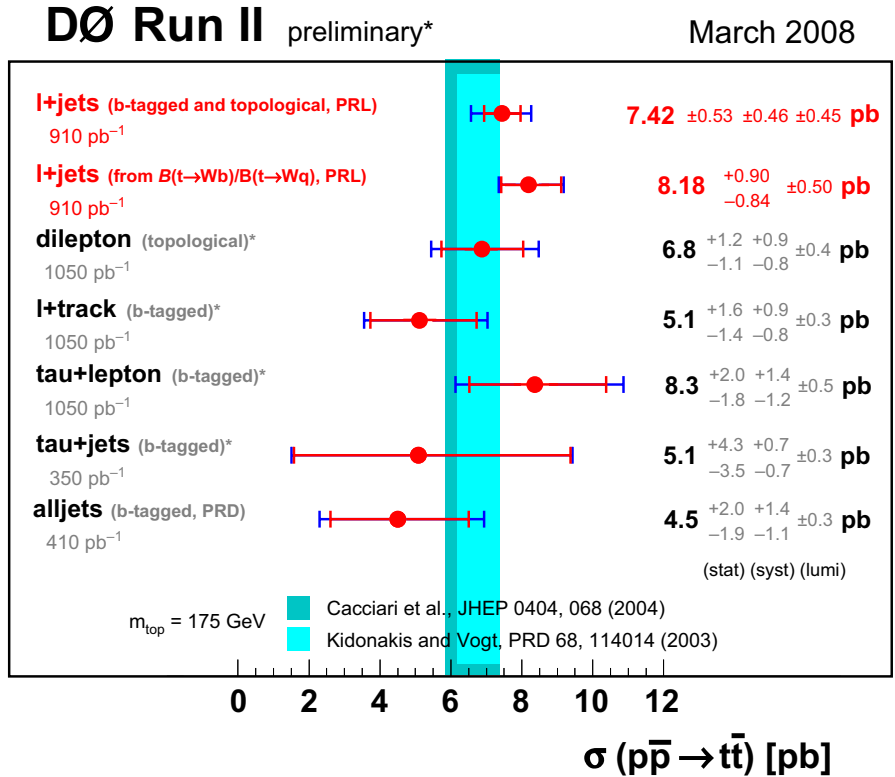}}
\caption{Summary of the D0 measurements of the $t\bar{t}$ production cross-section.
\label{fig:cs}
}
\end{figure}

The summary of the D0 $t\bar{t}$ cross-section measurements is shown on  fig.~\ref{fig:cs}.
All measurements are compatible with the standard model prediction.
The uncertainty on the measurement in lepton+jet channel  is now close to the theoretical  uncertainty and will
be improved with more statistics collected by D0 experiment. At the end of Run~II D0 plans to collect near 
8 fb$^{-1}$. With such integrated luminosity the statistical uncertainty in lepton+jets channel will be near 2\% 
and the total uncertainty will be limited by systematics and, in particular, by the luminosity uncertainty. 
Assuming the same relative systematics uncertainty 
as for the current measurement, one can expect to have near 8\% precision in the cross-section measurement at the end of Run~II.
Improving systematics uncertainty by a factor of two will decrease the total uncertainty to approximately 7\%.
The further improvement is not possible without changing the method of luminosity determination. For example, the possible approach 
could be to use $p\bar{p}\to Z$ events to normalize cross-section measurement. This approach has an advantage because all correlated 
systematics uncertainties will be canceled, but it also requires a careful investigation of the correlation between theoretical uncertainties 
of the $Z$  and $t\bar{t}$ production cross-sections calculation.

Another interesting measurement with the a statistics of Run~II is the cross-section ratio of
 dilepton and lepton+jet channels.
In this ratio the luminosity uncertainty and all correlated systematics uncertainties, 
\textit{e.g.} jet energy scale, jet and lepton ID identifications, will be partially canceled. 
The precision on this ratio will be limited by the statistical uncertainty in the dilepton channel and
could reach 5\% or even less. 

Cross-section measurement is also useful for the determination of the top quark mass using a theoretical 
dependence of the $t\bar{t}$ 
production cross-section on the  mass.
Such kind of measurements can not compete in precision with the direct measurements, but 
 they are much less dependent on the definition of the top mass parameter in MC generators.
The first results of such kind of measurement look quite  promising, see \textit{e.g.}~\cite{Abazov:2008gc}.

\begin{thebibliography}{0}

\bibitem{Kidonakis:2008mu}
 N.~Kidonakis and R.~Vogt,
 arXiv:hep-ph/0805.3844.

\bibitem{Cacciari:2008zb}
M.~Cacciari, S.~Frixione, M.~M.~Mangano, P.~Nason and G.~Ridolfi,
  arXiv:hep-ph/0804.2800.

\bibitem{Nadolsky:2008zw}
  P.~M.~Nadolsky {\it et al.},
  arXiv:hep-ph/0802.0007.

\bibitem{Tung:2006tb}
 W.~K.~Tung, H.~L.~Lai, A.~Belyaev, J.~Pumplin, D.~Stump and C.~P.~Yuan,
  JHEP {\bf 0702} (2007) 053
  arXiv:hep-ph/0611254

\bibitem{Abazov:2005pn}
  V.~M.~Abazov {\it et al.}  [D0 Collaboration],
  Nucl.\ Instrum.\ Meth.\  A {\bf 565} (2006) 463
  arXiv:physics/0507191.

\bibitem{Blazey:2000qt}
  G.~C.~Blazey {\it et al.},
  arXiv:hep-ex/0005012.


\bibitem{d0-tau}
D0 collaboration,
D0 note 5484-CONF, 
available at http://www-d0.fnal.gov/Run2Physics/WWW/documents/Run2Results.htm


\bibitem{Abazov:2008gc}
  V.~M.~Abazov {\it et al.}  [D0 Collaboration],
  Phys.\ Rev.\ Lett.\  {\bf 100} (2008) 192004
 arXiv:hep-ex/0803.2779.


\end{thebibliography}
\end{document}